# Causal Wave Mechanics and the Advent of Complexity.
# III. Universal Structure of Complexity

A.P. KIRILYUK*

Institute of Metal Physics, Kiev, Ukraine 252142

ABSTRACT. The universal dynamic uncertainty, discovered in Parts I and II of this series of papers for the case of Hamiltonian quantum systems, is further specified to reveal the hierarchical structure of levels of dynamically redundant 'realisations' which takes the form of the intrinsically probabilistic 'fundamental dynamical fractal' of a problem and determines fractal character of the observed quantities. This intrinsic fractality is obtained as a natural, causally derived property of dynamic behaviour of a system with interaction and the corresponding complete solution. Every branch of the fundamental dynamical fractal of a problem, as well as the probability of its emergence, can be obtained within the extended nonperturbative analysis of the main dynamic equation (Schrödinger equation in our case), contrary to basically restricted imitations of fractality within the canonical, single-valued approach. The results of the dynamical chaos analysis in Hamiltonian quantum systems, Parts I-III, are then subjected to discussion and generalisation. The physical origins of the dynamic uncertainty are analysed from various points of view. The basic consequences, involving essential extension of the conventional, unitary (= single-valued) quantum mechanics, are summarised. Finally, we emphasize the universal character of the emerging notions of dynamic multivaluedness (or redundance), causal randomness (or dynamic uncertainty), first-principle probability, (non)integrability, general solution, and physical complexity applicable to real dynamical systems of any kind.

NOTE ON NUMERATION OF ITEMS. We use the unified system of consecutive numbers for formulas, sections, and figures (but *not* for literature references) throughout the full work, Parts I-V. If a reference to an item is made outside its "home" part of the work, the Roman number of this home part is added to the consecutive number: 'Fig. 1(a)' and 'Fig. 1(a).I' refer to the same, uniquely defined figure, but in the second case we know in addition that it can be found in Part I of the work.

---

*Address for correspondence: Post Box 115, Kiev - 30, Ukraine 252030.
 E-mail address: kiril@metfiz.freenet.kiev.ua

## 4. Dynamic uncertainty and fractal structure of quantum chaos

In this section we consider more specific but basically important EP (effective potential) properties involved in quantum chaos and providing further insight into the dynamic complexity.

As the chaotic system dynamics is determined by the corresponding effective dynamical function, the properties of the latter should contain, in principle, all the intricacy of the chaotic behaviour. In the case of the effective potential in the modified Schrödinger equation, eqs. (5.I), (6.I), one may discern several related degrees of its structural dynamic complexity. The nonlocality and the dependence on the parameters are the most common EP properties (see also [1]), and they need not be directly related, in general, to the dynamic uncertainty. Nonetheless, these features represent already a kind of blurring and involvement reflecting the existence of another, explicitly hidden, degrees of freedom, and they can be influenced by chaotic dynamics.

The latter, however, manifests itself also in a more straightforward way by imparting the intrinsic dynamical indeterminacy to EP. This can be expressed directly by rewriting the formulas for EP, eqs. (6.I), for the $m$-th energy level of the $i$-th realisation; for example, eq. (6c.I) for the EP kernel is transformed into:

$$V^i_m(\mathbf{r}_\sigma,\mathbf{r}_\sigma') = \sum_{\mathbf{g}_\pi,n} \frac{V_{-\mathbf{g}_\pi}(\mathbf{r}_\sigma)V_{\mathbf{g}_\pi}(\mathbf{r}_\sigma')\psi^0_{\mathbf{g}_\pi n}(\mathbf{r}_\sigma)\psi^{0*}_{\mathbf{g}_\pi n}(\mathbf{r}_\sigma')}{\varepsilon^i_{\sigma m} - \varepsilon^0_{\mathbf{g}_\pi n} - \varepsilon_{\pi g_\pi} - 2\cos\alpha_{\mathbf{g}_\pi}\sqrt{(E-\varepsilon^i_{\sigma m})\varepsilon_{\pi g_\pi}}} \, ,$$

which means that operators $V_{\text{eff}}(\mathbf{r}_\sigma)$ and $\vartheta(\mathbf{r}_\sigma)$ acquire the same dependence on realisation and energy level, $V^{i\,m}_{\text{eff}}(\mathbf{r}_\sigma) = V_0(\mathbf{r}_\sigma) + \vartheta^i_m(\mathbf{r}_\sigma)$. In particular, the dependence on realisation number $i$ provides the 'chaotical' blurring of the function $V_{\text{eff}}(\mathbf{r}_\sigma)$, acting 'on the vertical axis'; it is superimposed on the nonchaotic blurring 'on the horizontal axis' due to the EP nonlocality. The characteristic magnitude of this dynamic EP splitting is determined by the chaotic regime in action (i. e. eventually by the parameters).

In the domain of global regularity, $K < K_c$ (see eqs. (20.II), (21.II), (30.II)) or $\omega_\pi > \omega_\sigma$, there are two types of the EP variations: the large, but relatively rare, quantum jumps and small 'trembling' passing to the stochastic layer in the semiclassical limit. As we have seen in the previous section, the amplitude of quantum jumps, determined by $\hbar\omega_\pi$, grows, in principle, infinitely with the distance to the classical border of global chaos, while the probability of their occurrence diminishes. Near the transition to global chaos this amplitude diminishes down to $\hbar\omega_\sigma = \Delta\varepsilon_\sigma$, and the jumps disappear as such. The "quantum stochastic layer" follows the inverse evolution with the distance to classical border. As is clear from Fig. 1(b).I, eqs. (14.I), and the accompanying analysis, far from this border, at $K \gg K_c$, 'trembling' gives the EP variations of the magnitude $\Delta V_{\text{eff}} \ll \Delta \varepsilon_\sigma$. With $K$ approaching $K_c$ they reach $\Delta \varepsilon_\sigma$ thus merging with the remnants of 'quantum jumps' and covering all the available motion domain. In the semiclassical limit this agrees qualitatively with the well-known results of classical description [2-4].

The 'smearing' manifestations of the fundamental dynamic uncertainty do not stop there. We are going to show now that the intricacy of chaotic dynamics leads inevitably to the existence of a next level of the EP structure complexity, still much more involved than, and superimposed on, those described above. Mathematically this type of 'superstructure' is eventually related to the same EP dependence on energy eigenvalues, $\varepsilon_\sigma$, in the form of resonance-induced singularities, eqs. (6.I), (14.I), that provides the fundamental multivaluedness (manifesting itself as the chaotic blurring above). To reveal it, we notice first that, by formally ascribing suitable values to $\varepsilon_\sigma$ in eqs. (6c,d.I), one can obtain practically every possible value of the EP kernel for arbitrary fixed values of other variables. Of course, it can never really occur in this way because we are confined to the self-consistent form of the modified Schrödinger equation, eq. (5.I). If one supposes, however, that the EP can take an arbitrary value at the respective restricted set of coordinate (and parameter) values, then this possibility may have been satisfied because the (topological) measure of this set provides another free parameter to adjust in order to satisfy the condition of self-consistency imposed by eq. (5.I). It implies also that the wave-function solutions for this equation possess similar property, and then they determine, entering the expression for EP, the quantum-mechanical probabilities, corresponding to that measure, for EP to take that value.

We can better see this possibility in our graphical representation of the modified Schrödinger equation, Fig. 1.I, eqs. (13.I), (14.I). Here, for example, arbitrary large EP values would correspond to the intersection point approach to the neighbouring asymptotes. To achieve this one should, for example, raise a small portion of the line $y = \varepsilon_{\sigma n} - \varepsilon^0_{\sigma n}$ by shifting it to the left, which corresponds to a decrease of $\varepsilon^0_{\sigma n}$. The latter quantity is determined, in its turn, by the free-motion energy, $h_{0n}$, in the corresponding state:

$$h_{0n} = \int_{s_\sigma} d\mathbf{r}_\sigma \, \psi^*_{0n}(\mathbf{r}_\sigma) h_0(\mathbf{r}_\sigma) \psi_{0n}(\mathbf{r}_\sigma) = -\frac{\hbar^2}{2m} \int_{s_\sigma} d\mathbf{r}_\sigma \, \psi^*_{0n}(\mathbf{r}_\sigma) \frac{\partial^2}{\partial \mathbf{r}^2_\sigma} \psi_{0n}(\mathbf{r}_\sigma) \, ,$$

where the last equality refers to the case of the simple configurational space representation (the other part of $\varepsilon^0_{\sigma n}$, the one due to the zero-order potential $V_0(\mathbf{r}_\sigma)$, can change to a much lesser degree). This free-motion energy can indeed be varied up to whatever high absolute values by changing the effective size (topological measure) of the coordinate set, mentioned above, supporting the state in question (in general, this follows from the uncertainty relation). It is important that this high absolute value can be accompanied by the negative sign of energy if the state is characterised by the 'localisation', which is a well-known and self-consistent property of the irregular states of this kind. This is the point which logically closes our preliminary demonstration of this new type of indeterminacy in the EP and the corresponding modified eigenvalue problem. In a similar manner one can obtain, with the respective finite probability, any value of EP and the other relevant quantities, the property which imparts quite a new quality to these objects and hence to the solution of a problem.

It is not difficult to deduce that an object strange enough to be compatible with the described EP properties should necessary be of fractal nature. Indeed, only fractal can be 'everywhere' and still possess the well-defined discrete structure. Of course, the above arguments can be regarded only as an evidence, rather natural and self-consistent, but not as a rigorous proof. To approach the latter, one should specify the above demonstration for particular tractable models using, most probably, computer calculations at least at some stages. We leave this special work for further investigations. However, already the qualitative demonstration above provides us with an important guiding line showing how the fractal can appear *directly* from the (modified) dynamic equations. In fact, it is a *solution*, for example, to the modified Schrödinger equation that in principle can be obtained and analysed, as we have seen, by analytical methods; this situation is similar to what we have for non-fractal special-function solutions of the ordinary linear differential equations. It is worthwhile to add that the involvement of fractals in chaos is a well-known fact (see e. g. [4-6]). Last time it obtains its quantum-mechanical extension by the evidence for the "fractal-like quasienergy spectrum", etc., in quantum chaotic systems (see e. g. articles [7] and the references therein). However, physical fractals typically appear in theory as a result of computer simulations, and their direct relation to the basic dynamical formalism remains obscure, which considerably limits the effective insight into the fundamentals of complex dynamics. In particular, common and distinct features of fractal-like structures and their manifestation at the level of observables for different types of chaos in real physical systems can be revealed only by exploring this direct relation between the deterministic dynamical equations and the fractal geometry. We present now a general scheme of such investigation within our formalism of FMDF (fundamental multivaluedness of dynamical functions), which is one of the possible ways of development of the qualitative arguments above.

First, we introduce explicitly the dependence of the 'zero-order energy', $\varepsilon^0_{\sigma n}$, on the topological measure, $\delta$, of the set supporting the respective eigenstate, $\psi_{0n}(\mathbf{r}_\sigma)$: $\varepsilon^0_{\sigma n} = \varepsilon^0_{\sigma n}(\delta)$. This measure also enters the definition of $V_{nn}(\varepsilon_{\sigma n})$, eq. (14.I), through the matrix elements $V^{nn'}_{\mathbf{g}_\pi}$, and thus our representation of the modified Schrödinger equation, eq. (13.I), should be written in the full form as

$$V_{nn}(\varepsilon_{\sigma n}, \delta) = \varepsilon_{\sigma n} - \varepsilon^0_{\sigma n}(\delta) \ .$$

For each $\delta$ we have a solution to this equation in the form of the set of realisations $\{\Re_i\}$, eq. (12.I), including the corresponding branches of the dynamical functions. Determining, for each of the realisations, the topological measure, $M_i$, of the obtained set of values of the dynamical function chosen (e. g. $V_{\text{eff}}^i(\mathbf{r}_\sigma)$, or $\{\varepsilon_{\sigma n}^i\}$, or $\rho_i(\mathbf{r})$), we obtain the scale-geometry relation, $M_i = M_i(\delta)$, which is one of the basic fractal characteristics and can be used for further study of its geometry [6,8]. We see that each of these fractals consists of $N_\Re$ *probabilistic* branches numbered by *i*, and its global geometry is characterised by their full ensemble, $M(\delta, \mathcal{D}) = \{M_i(\delta)\}$. It is very important that this fundamental geometrical relation inherits the dependence on the dynamics $\mathcal{D}$, eq. (17.II). It represents, in the explicit form, a way for the *full geometrisation of dynamics*, now also for chaotic (including quantum) systems.

Because of the fundamental role of the effective dynamical function in chaotical system dynamics, the corresponding fractal bears a special significance. We call it the *(fundamental) dynamical fractal*, the term representing, in fact, a synonym of the effective dynamical function that emphasises its fractal character and encoding of the complete chaotic dynamics in its geometry. Already the general qualitative features of this geometry, outlined above, demonstrate the high level of intricacy in the structure of chaos. The latter manifests itself geometrically as the nonlocal multi-branch fundamental dynamical (multi)fractal. This specifies also the involvement of the fundamental multivaluedness: we see now that the details discussed in previous sections concern the non-fractal, 'solid' part of the branches which are surrounded by their 'semi-transparent' fractal 'foliage'. It is important to emphasize, however, that the fractal structure of EP and other measured quantities can also be considered as a result of additional splitting of the obtained realisations of the 'first generation', proceeding through the same FMDF mechanism. This additional splitting starts, mathematically, at the level of auxiliary equations (eqs. (4a.I)) which provide an *approximately* closed solution *only* at the considered (first) level of the fractal hierarchy of dynamical multivaluedness, while finer dependence of the quantities participating in the basic FMDF analysis (like $\varepsilon_{\sigma n}^0$ or $V_{\mathbf{g}_\pi}^{nn'}$) on the solutions to be found gives an infinite chain of (causally probabilistic) fractal branching. This approach confirms the above 'geometrical' consideration and provides another aspect, and way of investigation, of the fractal structure of complexity (it shows, in particular, why any usual, single-valued approach cannot provide the causally complete, dynamical version of fractal structures).

Fractal structure of dynamical functions does not need to disappear in a regular regime of a chaotic system. Our graphical method of analysis shows, however, that the most significant, relatively dense, parts of the fractal web are concentrated around 'solid' parts of the function, in their vicinity limited to the dynamic uncertainty (the "quantum stochastic layer", see above in this section and section 3.II). This implies that the fractal web tends to 'shrink' around the deterministic carcass with growing regularity in the system, even though some looser part of it seems to be always present almost everywhere, smoothly disappearing only asymptotically, similarly to the stochastic layer.

Further studies are necessary to specify the role of random and deterministic elements in the fractal structure of chaos. Here we emphasize, however, the difference between the multi-branch coarse-grained probabilistic structure of the fundamental dynamic fractal, providing the main irreducible source of randomness and dynamic complexity (section 6), and the involvement of the dynamic fractal geometry on the fine scale leading to specific character of the system dynamics even within the same globally deterministic realisation (the involved fractal-like form of the dynamical patterns, 'omnipresent' and still inhomogeneous, etc.). At the same time, these two aspects of the complex (quantum) dynamics are closely related already by their origin, the effective nonlinear self-interaction of the complex system (section 6) being represented by the particular structure of the effective dynamical function in the modified form of the main dynamic equation.

This relation will manifest itself also at the level of the measured quantities and the observed physical effects. In particular, the fractal structure of realisations can considerably attenuate the partial quantum suppression of chaos discussed in sections 2.3.II, 3.II. Indeed, the latter is due, generically, to the finite realisation separation for quantum chaotic systems (section 2.2.I). The existence of the fractal web means, however, that the system can find itself, with a finite probability, somewhere in between the realisations, and it can even traverse the whole 'distance' separating the neighbouring realisations by an involved diffusion-like evolution on the fractal filling that space. This effective disappearance of the noise magnitude threshold for realisation change can be described also as the high sensitivity of a chaotic system to small noises, a well-known property of the classical chaotic systems [4]. Now this property is extended to quantum chaotic systems, and in general to complex systems with the finite realisation separation. Note, however, that the partial suppression of chaos in such systems will be preserved, though in a reduced form, because the diffusion-like wandering on a fractal is not equivalent to the zero-threshold direct realisation change.

In a somewhat similar fashion, the fractal structure of realisations should lead to reduction of another obstacles to the system motion, the (effective) potential barriers. We have seen, in the previous sections, that in each of the realisations, separated but, as we have just shown, partially connected between them, the system moves in a corresponding branch of the effective potential, generically resembling the zero-order potential and in particular containing potential barriers. At the end of the previous section we discussed one specific way for a chaotic system to traverse such a barrier, called chaotic tunnelling and related to variation of the effective potential barrier height with realisation change. Now we can propose a supplementary mechanism of chaotic tunnelling consisting in penetration through a barrier by a diffusion-like fractal wandering. It is possible due to the fractal structure of the effective barrier itself: it is a part of the fundamental dynamical fractal. As a result, the EP barriers are not solid, each of them contains a more or less dense fractal net of 'holes' forming a kind of the involved labyrinth. This means that such barriers, even when they are wide enough, cannot confine a system infinitely, and each 'bound state' has the corresponding finite width and lifetime. Physically this mechanism of chaotic tunnelling in the modified, effective description is equivalent to the system 'going round' the prototype barrier in the ordinary description containing more dimensions (degrees of freedom).

A question can be posed about the relation of all those types of chaotic tunnelling to the ordinary quantum tunnelling appealing to the exponentially small "wave-function penetration" within the barrier, physically taken rather as axiom. The evident conclusion that this ordinary tunnelling is just added to the described chaotic components seems to be rather superficial and incomplete. Instead, we may argue, based on the results above and also those of parts IV,V, that even the 'ordinary' quantum tunnelling can eventually be largely reduced to chaotic tunnelling, taking into account that almost any real quantum system is chaotic, to some degree. The full development of the corresponding theory should include the intimate involvement of chaos with the foundations of quantum mechanics (parts IV,V) and deserves a special consideration. Nevertheless, some basic lines of this general theory stem directly from the above dynamical uncertainty implication in quantum tunnelling, demonstrating its potential fundamental importance.

Concluding this section, we emphasize that the described natural emergence of fractal structures in complex system dynamics seems to possess the same universality as the method of FMDF in general (section 6) giving the real hope to unite the known (and unknown) manifestations of fractals in various chaotic systems under the unique scheme of the complete geometrisation of dynamics.

In particular, it is difficult not to see the direct relation between the fundamental dynamical fractal introduced above and the well-known fractal objects of classical chaotic dynamics, such as Cantor-tori and strange attractors. Based on the general character of our method and its main results (section 6), we may argue that all such objects are particular manifestations of the fundamental dynamical fractal for the corresponding types of dynamical systems (e. g. for the Hamiltonian classical systems in the case of Cantor-tori and for dissipative classical systems in the case of strange attractors). The crucial advantage of our representation of fractal dynamics is that those basic fractal-like objects are explicitly obtained as the *general solutions* of the relevant dynamical equations in the modified form, the method applied being quite universal and allowing, in principle, for analytical derivation of the fractal properties in relation to the dynamical system parameters (this is precisely the sense of the full geometrisation of the chaotic dynamics). Moreover, one can see the universal physical origin of the fractal structure of chaos as a manifestation of the effective dynamical instability, the latter being also at the origin of the causal randomness (section 6), another basic feature of chaos. The definite confirmation of these statements will demand the direct application of the described methods to the corresponding dynamical systems, which provides a subject for further studies that are now in progress (see e-print physics/9806002).

## 5. Discussion

The concept of chaos suggested above and underlying all the subsequent detailed calculations is centred around the concept of the fundamental dynamic uncertainty obtained in the form of FMDF which reveals the existence of many (more than one, at least) complete solutions of the modified Schrödinger equation, eq. (5.I). The effective advantage of the latter for chaotic systems, as compared to the ordinary form, is taken as axiom supported by the results of the previous sections. It should eventually be verified also by comparison of the corresponding predictions specified for particular physical systems (e. g. [17]) with the experimental results for them. At the same time, the concept of FMDF and the formalism proposed may need some additional general explanations.

First of all, it is important to note that the 'unexpected' plurality of solutions itself does not contradict any existent basic principles and theorems. In particular, one may recall that the ordinary uniqueness theorems limit the number of solutions of the Schrödinger equation to one if the participating functions possess certain 'normal' properties. The apparent disagreement between this basic restriction and the postulate of fundamental multivaluedness disappears in the obvious way once we notice that the mentioned theorems never assume, directly or indirectly, either the multivaluedness of the potential, or its explicit dependence (which is a singular multi-branch one in our case) on the eigenvalues to be determined. We have seen, however, that it is just the last property which leads, in a self-consistent manner, to the additional plurality of the EP branches and the corresponding solutions, eq. (12.I). In fact, the uniqueness remains, simply now it means that *each realisation* corresponds to only one *branch* of EP and only one complete set of eigenfunctions with their eigenvalues. Of course, this statement needs a rigorous mathematical consideration starting from the modified Schrödinger equation (5.I), which is out of scope of the present paper. The general explanation of the apparent contradiction between the new and old formulations of dynamical equations, the respective mathematical schemes, the perturbative and non-perturbative treatments, and the computer simulation results (see below) can be reduced thus to the relative self-consistency of each of the approaches, which demand special efforts for passing from the old paradigm to the new one.

The physical meaning of the fundamental multivaluedness can be illustrated also in a way similar to that for the ordinary multivaluedness of mathematical functions describing measurable physical quantities. Namely, the transition from the ordinary to the modified form of the Schrödinger equation is analogous to that from, say, $f(x) = +\sqrt{x}$ to $[f(x)]^2 = x^2$ with the solutions $f(x) = \pm\sqrt{x}$. The fact that we may consider the first, restricted, form of a function to be sufficient and the second, multivalued, one to be excessive depends on physical arguments; for example, there may be some reasons to regard only positive solutions to be 'physical'. However, if we discover that negative solutions can also have physical meaning, then we accept the second, larger, form which *includes* the first one. In the story with quantum chaos, the restricted, or 'ordinary', form of the Schrödinger equation with a single-valued solution corresponds well to the integrable problems with regular dynamics which are, in fact, rather special and rare, as is clear today. Once we try to apply it to more general situations involving chaotic behaviour, we notice a striking contradiction, the famous 'pathological' regularity of *this form* of quantum mechanics. Then it would be relevant to try the extended one, that of the modified Schrödinger equation, giving multivaluedness which, as we have shown, is sufficient to give a self-consistent quantum chaos description passing smoothly (though not trivially !) to the well-established classical results. It is important, that this scheme does not directly interfere with the conventional interpretations of quantum mechanics and the fundamentals of its formalism: the extended quantum dynamics is 'derived' from the ordinary representation and includes, in a modified form, the solution that gives regular dynamics in the absence of chaos. In the latter case this solution constitutes a single realisation, while the others are suppressed due to quantum effects (see also their discussion below) or, as we have seen, are asymptotically indiscernible from it. When the pronounced chaos appears, the other components of the 'fundamental multiplet' are added to the existing one, and they form, all together, the complete solution of a problem. One may say, thus, that the introduced dynamic uncertainty does not directly depend on the quantum-mechanic uncertainty. This does not exclude the reverse and below, in part IV, we show that fundamental quantum indeterminacy can be understood as a particular manifestation of the same dynamic uncertainty described by the FMDF. Note that this natural including of the modifications to the existing single-valued forms is profoundly related to a particular distinction of the fundamental dynamic multivaluedness from the ordinary multivaluedness mentioned above: the former has a pronounced dynamic character, the number of branches is determined from the dynamical equation and can vary depending on parameters; this explains also the intricate fractal ramification of the obtained branches.

It might be useful to summarise, in the explicit form, the described 'gentle' modification of quantum formalism for the analysed case of the multi-dimensional Schrödinger equation with the potential periodic in one of the dimensions, eq. (1.I). In order to simplify the expressions (but without any real limitations, see section 2.1.I for the full version) consider the case of a one-dimensional perturbation, $\mathbf{r}_\pi \equiv z$, with the period $d_z$ (i. e. in the formulas above $\mathbf{r} = \{\mathbf{r}_\sigma, z\}$, $\mathbf{g}_\pi \equiv g_z = 2\pi g/d_z$, where $g$ is a non-zero integer). Then instead of the Schrödinger equation in the ordinary form,

$$\left(-\frac{\hbar^2}{2m}\right)\frac{\partial^2}{\partial \mathbf{r}_\sigma^2}\Psi(\mathbf{r}_\sigma,z) + \left(-\frac{\hbar^2}{2m}\right)\frac{\partial^2}{\partial z^2}\Psi(\mathbf{r}_\sigma,z) + V(\mathbf{r}_\sigma,z)\Psi(\mathbf{r}_\sigma,z) = E\Psi(\mathbf{r}_\sigma,z), \qquad (32)$$

one should use the modified form:

$$\left(-\frac{\hbar^2}{2m}\right)\frac{\partial^2}{\partial \mathbf{r}_\sigma^2}\psi(\mathbf{r}_\sigma) + V_0(\mathbf{r}_\sigma)\psi(\mathbf{r}_\sigma) +$$

$$+ \sum_{g,n} \frac{V_{-g}(\mathbf{r}_\sigma)\psi^0_{gn}(\mathbf{r}_\sigma)\int_{s_\sigma} d\mathbf{r}_\sigma'\, \psi^{0*}_{gn}(\mathbf{r}_\sigma')V_g(\mathbf{r}_\sigma')\psi(\mathbf{r}_\sigma')}{\varepsilon_\sigma - \varepsilon^0_{gn} - \varepsilon_{\pi g} - 2\pi\hbar^2 K_z g/m d_z} = \varepsilon_\sigma \psi(\mathbf{r}_\sigma) , \quad (33a)$$

the solutions of this equation used to obtain the total wave function $\Psi(\mathbf{r})$ with the help of the following relation:

$$\Psi(\mathbf{r}_\sigma, z) = \exp(izK_z) \times$$

$$\times \left( \psi(\mathbf{r}_\sigma) + \sum_{g,n} \frac{\exp(2\pi i g z/d_z)\psi^0_{gn}(\mathbf{r}_\sigma)\int_{s_\sigma} d\mathbf{r}_\sigma'\, \psi^{0*}_{gn}(\mathbf{r}_\sigma')V_g(\mathbf{r}_\sigma')\psi(\mathbf{r}_\sigma')}{\varepsilon_\sigma - \varepsilon^0_{gn} - \varepsilon_{\pi g} - 2\pi\hbar^2 K_z g/m d_z} \right), \quad (33b)$$

where

$$V(\mathbf{r}) = V_0(\mathbf{r}_\sigma) + \sum_g V_g(\mathbf{r}_\sigma)\exp(2\pi i g z/d_z) ,$$

$K_z^2 = 2m(E - \varepsilon_\sigma)/\hbar^2$, and $\{\varepsilon^0_{gn}\}$, $\{\psi^0_{gn}(\mathbf{r}_\sigma)\}$ are the eigen-solutions for the auxiliary system of equations (4.I). We refer to the corresponding definitions of other participating quantities given in section 2.1.I, with the evident substitutions involving $z$ and $g_z$. Note that in the general case the above expansion of the total potential, $V(\mathbf{r})$, is performed not necessary in plane waves but in any appropriate complete orthonormal set of functions determined mostly by the symmetry considerations. Correspondingly, the choice of the 'regular' part, $V_0(\mathbf{r}_\sigma)$, within the total potential may not be unique (the only limitation, rather of practical character, is that the unperturbed Hamiltonian, $H_0(\mathbf{r}_\sigma)$, should be integrable), and then it is determined by reasons of convenience (symmetry, etc.). Although the modified form, eqs. (33), is obtained from the ordinary one, eq. (32), by the straightforward, in principle, algebraic transformations described in section 2.1.I (the compatibility of the two forms can also be verified directly), it is evidently much more involved, even by its external presentation. As we have seen in the preceding sections, it is this intricacy that represents eventually the unavoidable (and probably minimum) 'payment' for revealing the true quantum chaos.

One can now specify and understand the 'mistake' of the conventional application of the ordinary form of the Schrödinger equation to the description of *any* quantum system. First of all, we see that the basic problem with the case chaos does not need to be attributed to the very foundations of quantum mechanics including the probabilistic wave paradigm, correspondence principle, etc., as it is implied in [10]; neither needs it be reduced to the intricacy of the semiclassical limit [11]. The conceptual and, eventually, practical difficulties with *general* quantum paradigm addressed to by Einstein, de Broglie and others certainly exist, but what we have shown here is that it is another story which is not necessarily *directly* involved in the problem of quantum chaos (this does not preclude the eventual implication of non-linearity and chaos in this story, see parts IV, V). As far as chaos is concerned, the 'mistake' was in a tacit *straightforward* extension of the ordinary form of the Schrödinger equation, eqs. (1.I), (32), well established and verified, in fact, practically only for regular cases (see also the next paragraph), to arbitrary situations. Now, a posteriori, one may invoke evident, though not rigorous, physical considerations showing that this extension has been rather doubtful, even from a general point of view. Indeed, what real physical sense can one imply in the conventional form of the Schrödinger equation (see eq. (32)) written, for example, with the three-dimensional periodic potential, $V(\mathbf{r}) = V(x,y,z)$, presented by an ordinary single-valued function, i. e. in fact, by a number? The absence of a good answer to this question becomes more evident if we first pose it for a regular case, say, that with $V(\mathbf{r}) \equiv V(x)$. For in *this* case the answer is well-defined and can be expressed in many forms. Mathematically, a solution is reduced, in principle, to quadratures, i. e. a number gives another number. Physically, one may speak, for example, about multiple scattering of waves from the opposite sides of the potential well(s) $V(x)$ giving finally, due to interference, only a fixed number of well-defined standing waves which are 'confined' by the well(s). Now with an arbitrary three-dimensional potential, how can the wave be ever 'confined' between two potential surfaces if there is very often an 'issue' by another dimension? And what is this scattering and interference, if the slightest change of the conditions of each scattering (or an interference path) leads to an unpredictably large modification of the result, while the interfering partner always tends to 'run away' to infinity through multiple 'roundabout' ways? These considerations, though not rigorously specified, seem to leave the definite impression that, for a three-dimensional problem, the form (32) is nothing but a symbolic script which can serve to fix the conditions of a problem, but certainly, in principle not to provide any meaningful solution. Despite this pessimistic conclusion, one could try to overcome the difficulty in a logically straightforward way: let us start with a basically non-contradictory part that we understand well (i. e. $V(x)$) considering the rest (i. e. $V(x,y,z) - V(x)$) as perturbation (not necessarily small). This leads directly to the formulation (33) which appears to be a *new, extended* presentation of the Schrödinger equation because of the fundamental multivaluedness that it naturally reveals, and the courage needed to accept it, is recompensed by the opening possibility of the non-contradictory quantum treatment of however complex chaotic systems. This *a posteriori* construction of quantum mechanics possible only in the apocalyptic epoch of dynamical indeterminism, paradoxically permits one to delay, once more, the anticipated fall [10] of the other indeterminism seemed to be much more mysterious. This may leave a disturbing impression of the Creator playing simultaneously several different games of chance. We believe, however, that the remaining 'conventional' incompleteness of quantum mechanics can be eventually removed in favour of the universal dynamical complexity (see parts IV,V for the details), and then the suggested modification of quantum formalism gives a support for *this* way of development of quantum paradigm.

A particular question arises about the exact range of applicability of the ordinary form of the Schrödinger equation. In this case our description gives a rather definite and 'convenient' answer: the only domain of the rigorous validity of the conventional form of the Schrödinger equation for non-separable problems[#] is that of the complete absence of chaos, which is the regime of quantum suppression of chaos outside the quantum border of chaos, $E < E_q$ (see eq. (22.II)). The convenience consists in the fact that, as we have shown in section 3.II, it is just this regime that is realised for an elementary quantum system in the ground state, the latter forming the basis of the conventional 'regular' quantum mechanics.[*] Contrary to this, higher energy states always involve *some* chaoticity and therefore, strictly speaking, the modified form of quantum-mechanical formalism is necessary for their description. This concerns also partial regularity beyond the classical border of chaos and, most important, the *whole time-dependent case*. Indeed, for the latter, as we have seen, the suppression of chaos is always partial because the total quantum suppression does not exist. Therefore the time-dependent quantum formalism to be exact always needs to be formulated in the modified version. All these rigorous limitations do not preclude, of course, the possibility of good *approximate* validity of the conventional description for many cases when, as we have seen, chaos is relatively weak. This explains visible 'absence' of the known experimental manifestations of quantum chaos which may well be hidden somewhere in the "linewidths" attributed to numerous other reasons. This is related also to the basic difficulty of experimental detection of quantum chaos manifestations: the unavoidable quantum uncertainties and small stochastic influences 'smear' every observed pattern even without any chaos which, when it exists, should be then somehow separated from them. It is not easy, however, to separate a noise from its noisy background! Of course, this problem can finally be resolved by special theoretical and experimental investigations of linewidths and other chaos-sensitive quantities (a good opportunity to detect quantum chaos experimentally is presented, for example, by the specific regime of chaotic quantum jumps, section 3.II).

---

[#] I. e. those which cannot be reduced to a number of effectively one-dimensional problems. They include practically all real systems with very few exceptions. Among the most popular cases only the hydrogen atom seems to be a good model of a regular quantum system, provided that all excitation fields are excluded, as well as more complex secondary interactions (spin-orbit, etc.). Numerous 'one-dimensional' problems analysed within the ordinary quantum mechanics are nothing but approximations that can work rather well (though not without 'surprises') only in limited parameter ranges.

[*] See, however, the note on the possible existence of the chaotic ground states in section 3.II; if this possibility, apparently quite reasonable, is confirmed, then even this application of the conventional formalism to ground state description may be incomplete.

The results of the previous sections provide also a transparent *physical* explanation for the plurality of realisations and the ensuing origin of quantum chaos. Indeed, in terms of the time-independent case, the dynamics of a system can be considered as the interplay between its two related parts, the 'unperturbed' one corresponding to coordinates $\mathbf{r}_\sigma$ coupled to the zero-order Hamiltonian $H_0(\mathbf{r}_\sigma)$, and the 'perturbing' one represented respectively by $\mathbf{r}_\pi$ and $H_\mathrm{p}(\mathbf{r}_\sigma,\mathbf{r}_\pi)$ (cf. eq. (1.I)). In particular, there is a permanent energy exchange between the two groups of degrees of freedom, $\mathbf{r}_\sigma$ and $\mathbf{r}_\pi$, eventually described in our formalism (see the transition from the system (2.I) to eq. (5.I)) by the denominators in the expression for EP, eq. (6c.I). The corresponding energy balance, consistent with the unchanged total energy, $E$, is achieved when one of these denominators equals to zero (this corresponds to a resonance, in terms of classical mechanics). This condition, as we have seen in the previous section, means also the existence of the related asymptotes in our graphical representation and thus, eventually, of another realisation of a problem. Then the existence of more than one realisation can be expressed in simple physical terms just as the *nonuniqueness* of the way to satisfy this energy balance condition. In other words, the total energy division between the two groups of degrees of freedom can be done in many different ways. In a chaotic system (which is, in fact, just a system with a non-trivial configuration) the number of these ways easily exceeds the one within a single realisation, and then other realisations necessarily appear. This interpretation provides equally clear explanation for the existence of the quantum border of chaos, eq. (22.II). Indeed, in quantum mechanics the mentioned energy exchange can proceed only in discrete quantities, and it is evident that if the total energy is less than the minimum of these quantities then *any* exchange is strictly prohibited. That is why we obtain the *complete* suppression of chaos below the quantum border, the fact of fundamental importance for the stability of atoms and nuclei as it was shown in the previous section. It is also clear why the quantum border does not exist in the time-dependent case: the total energy is not fixed, and the corresponding limitation cannot be imposed. Nonetheless, the energy exchange exists in this case also, and the corresponding interpretation of chaos by the fundamental multivaluedness remains valid.

This physical explanation of the origins of quantum chaos shows that, in fact, it is reduced to a special kind of instability: there are more than one equivalent possibilities (represented by realisations) for a system, and it 'does not know' which one to choose. This permanent and unavoidable 'hesitation' of a system is manifested as the inherent stochasticity of dynamics. This takes us back to the known concept of instability as the basic origin of chaos in classical mechanics. From one hand, we can make an important conclusion that chaos is *always* based on instability, both in quantum and classical mechanics. From the other hand, we see that in quantum case our approach provides an *explanation for the instability itself* based on a *more fundamental* and *physically simple* concept of multivaluedness which naturally appears also in the corresponding mathematical formalism. All this permits us to suppose that an analogous interpretation of the notion of instability can be found also for the classical description of Hamiltonian chaos and eventually for other types of chaos extending considerably their understanding (see also section 6).

Note that once obtained and confirmed the postulate of multivaluedness seems to be much more natural, even from the formal point of view, and consistent with the existing reality, than the conventional 'single-valued' paradigm. Indeed, the acceptance of multi-valued functions as the basis for dynamics is evidently a much more general choice than that of single-valued functions, and thus, a posteriori, it would be preferable already because of this formally wider nature (especially taking into account the profound dynamical character of the multivaluedness discussed above). If we accept this wider choice then we are obliged to explain not the complex phenomena like chaos and the multiplicity of forms which become quite natural, but rather relatively rare and 'pathological' exceptions of regular dynamics. This logic of our approach seems to be much closer to the general physical picture of the world as we see it now. As to the explanation of the occasional regularity, we have seen that it is always related to the reduction, complete or approximate, of the number of realisations down to one (apart from the specific effect of quantum discreteness on the transitions between the realisations which favours quasi-regularity and not the real regularity). Physically this reduction can be attributed either to decreasing instability due to the lower choice of possible energy redistributions near the quantum border of chaos, or to the growing similarity of realisations outside the classical border.

It is important to emphasize also that many particular features of chaotic behaviour in quantum systems remind qualitatively the results obtained for complex system of apparently quite different origin. For example, as concerns the parameter dependence of system behaviour, the appearance of new dynamical regimes around some parameter values, '*bifurcation* points', due to the 'ramification' of the system 'state' is well-known in self-organisation theories (synergetics), also as a manifestation of the non-equilibrium phase transitions, in catastrophe theory, in various types of dynamical chaos (e. g. turbulence), etc. Moreover, the nature itself of our realisations for a chaotic system resembles much the properties of the self-organised states introduced in synergetics and "dissipative system" analysis [12] with their generalised "order parameters", "mode enslavement", etc. Now one can apply the same general terms of bifurcation theory and synergetics to quantum chaos. In section 6 we show that this analogy has a very profound basis: it is a manifestation of the same unique origin of complexity in various dynamical systems, and the concept of the fundamental dynamic uncertainty, proposed above for quantum Hamiltonian systems, can be extended to the universal understanding, and a method of practical analysis, of the complex dynamic behaviour of any particular origin. This can be used to better understand, and to put into a self-consistent general scheme, the existing various types, notions, and concepts of now vaguely defined 'complex' behaviour of arbitrary self-organised chaotic systems.

Emphasizing the basic role of the fundamental multivaluedness, it is important to note, however, that the detailed difference between regular and chaotic dynamics is yet more complex and contains many other features. In particular, each of the realisations within the chaotic regime seems to differ qualitatively from the (quasi-)unique realisation within the regularity. It means that, whereas FMDF serves the indispensable source of unpredictability and hence complexity (see below) for quantum chaos, the latter is characterised *also* by a number of other specific properties concerning *each* solution-realisation which are absent, or differ significantly, in the unique realisation within the regular dynamics. These specific properties were extensively studied within the conventional approaches to quantum chaos (see e. g. [13]) and are known as signatures of chaos (cf. section 1.I). The most popular among them are the specific statistical characteristics of energy spectrum like the distribution of energy-level spacing, and the peculiar coordinate dependences of PDD (probability density distribution) for chaotic wave functions. We suppose that most of these properties known from experiment or computer calculations does not contradict our approach and the discovered FMDF; they can be reproduced by the same, or another, means within this approach being effectively added to the features already revealed within the fundamental multivaluedness. It does not preclude, however, their modification in details and in explanation of their origins that worth a separate investigation. One may mention, for example, the straightforward effect of energy-level attraction within the domain of global regularity due to the existence of many similar realisations, see Fig. 1(b).I and the explanations in section 3.II. From the other hand, a typical energy-level separation within the global chaos regime seems to be close to $2(\varepsilon_{\pi}g_{\pi 0}E)^{1/2}$ (Fig. 1(a).I) evidently favouring the energy-level repulsion. We restrict ourselves to these particular illustrations with the conclusion that the existence of signatures of (quantum) chaos seems to be a good complement to the fundamental multivaluedness demonstrating the high level of intricacy of dynamical chaos (another evidence of the latter is provided by the unravelled fractal structure of the chaotic quantum dynamics superimposed on the fundamental multivaluedness, section 4).

Another part of the problem of relation between our approach and the existing results on quantum chaos has an apparently simple form. The question is: why the discovered additional plurality of solutions does not appear in the numerous studies of chaotic quantum systems by other methods and especially in computer calculations often based on rather elaborated approximations? The answer is: it is just because one always uses the ordinary form of the Schrödinger equation (or other equivalent formalism) which somehow hides the real plurality of solutions (but not the signatures of chaos! - see the preceding paragraph). However, already this mysterious hiding provokes further questions returning us to the physical meaning of multivaluedness. Indeed, if one would be able to implement a really perfect simulation scheme for, say, three-dimensional Schrödinger equation with periodic potential, but in the ordinary form, then is there a chance that the obtained computer 'virtual reality' naturally reproduces the fundamental multivaluedness of realisations and the random transitions between them? If the answer is no, then it would mean that to obtain the multivaluedness and real chaos it is absolutely necessary to use a special form of the dynamical equation (the modified Schrödinger equation in our case), whatever the scheme of the simulations. Although the striking difference between the conventional results of computer simulations and the predictions of our approach suggests rather this negative answer, we prefer not to insist on it until the accomplishment of the detailed study on the appearance of multivaluedness in simulations, which is one of the subjects for further research. Indeed, much may depend, in the actually unknown fashion, on the complexity of a problem (e. g. its effective dimensionality) and on the details of the algorithm concerning, in particular, the approximations used. Thus, for example, the authors of ref. [14] obtained true chaos in their simulations of a quantum many-body system undergoing multiple resonant tunnelling; in general, many-body systems seem to exhibit the true chaos more readily, even in quantum mechanics. One should not forget also that the results of simulations, especially at the level of observables, often leave a great freedom for their possible detailed interpretations. Finally, it is important to emphasize that all these considerations on the relation with the results of the conventional calculation techniques do not influence directly the main conclusions of our method with its own scheme of calculations providing, as we have shown, the real dynamical chaos in agreement with the well-established results of classical mechanics.

As concerns the possibility of the fundamental multivaluedness appearance within other analytical approaches to quantum chaos description, it seems to be generally probable, but not easily realisable; this is eventually due to the same particularities of the relation chaos - quantum mechanics that determine the well-known difficulties of the 'dynamical randomisation' of quantum mechanics (and of any other 'nonlocal' theory, see section 10.V). The suggested problem reformulation within the generalised method of the effective dynamical functions may prove to be objectively a rather unique opportunity for resolving the contradiction between the quantum-mechanical indeterminacy and the stochastic uncertainty of dynamical chaos. This approach gives an opportunity, rare for a quantum description, to deal with an *observable* quantity, EP, manifesting the chaotic uncertainty via FMDF, but entering *also* the *main dynamical relation*, the Schrödinger equation, for the *unobservable* wave function with its quantum indeterminacy. Note that previously an attempt was made to use the EP formalism, though in the operator form, for quantum chaos description [15,16], but the subsequent analysis went in the direction quite different from that of the present work and not touching the idea of the problem splitting into multiple realisations. Therefore the results obtained in [15,16] have eventually very little in common with those of the present paper. Among other general formalisms that can be more susceptible of revealing an analogue of the fundamental multivaluedness, we mention the Feynman path integral and the density matrix.

# 6. Dynamic multivaluedness, (non)integrability, completeness and complexity of dynamical systems: a unified approach

Summarising the results of the proposed description of quantum chaos by the generalised EP method and the derivative concept of the fundamental dynamic uncertainty, it is important to note that their essence does not depend on the form of the particular systems or models considered. Thus one can easily extend the presented description of chaos induced in a Hamiltonian system by the addition of extra degrees of freedom (dimensions) to the case of chaos induced by the change of symmetry of the potential. The basic mechanism of the problem splitting into a set of realisations, depending on parameters, with the 'spontaneous transitions' between them remains always valid for such an extension, as well as the method of the detailed analysis of the realisations, while the resulting particular types of chaotic behaviour and the conditions for the transitions between them are determined by the details of the splitting and may vary for different system types.

Moreover, we may suppose that this generalisation can be extended to unite under the one universal approach all major cases of dynamical chaos now practically separated: classical conservative systems, nonlinear dynamical systems, quantum and classical dissipative dynamical systems, and finally distributed nonlinear systems, presenting the highest level of complex behaviour. This assumption should be proved by further practical developments of the method, but it can be partially justified already now by a number of rather 'persistent' considerations such as:

the universality of the formalism of FMDF (for example, practically by a simple change of notations it is easily generalised to any problem with linear "ordinary" formalism);

the consistence of the main qualitative predictions of the method with the observed patterns of chaotic phenomena in the form of the permanent irregular change-over of quasi-regular motion regimes (or structures) belonging to a fixed discrete (or, sometimes, quasi-continuous) set;

the flexibility of our method and the diversity of dynamic regimes it predicts already in the general terms including, in particular, transitions chaos-regularity in the parameter dependence, simply expressed and clearly classified cases of 'strong' and 'weak' chaos, of partial and total suppression of chaos, intermittence of chaotic regimes, e. g. in the domain of global regularity for periodically perturbed systems, etc.;

the natural involvement of fractals in our approach which appear as a universal form of solutions of the modified dynamical equations completing the geometrisation of dynamics;

the well-supported non-controversial explication of the existence of true chaos even for 'hard' quantum case and the attractiveness of a unique general basis for any kind of dynamical chaos (see also the end of section 10.V).

Of course, the full realisation of this unification need not always be straightforward and will demand certain well-defined choices like, for example, that of the proper starting formulation of a problem. In return one obtains the unification not only for different fields of physics and types of chaos, but also for apparently different (and even sometimes opposed) general insights on complex systems, like those of chaos, self-organisation, nonequilibrium statistical physics, and catastrophe theory. It is not difficult to see that in the frame of the concept of dynamic multivaluedness all of them become nothing but particular limiting cases, or presentations, of the same multivalued reality which can be explicitly obtained one from another, for example, by variation of parameters. Whereas the implementation of this program and verification of the underlying suppositions demand a series of separate detailed investigations, we consider it to be useful to outline here in this concise fashion the anticipated universal meaning of the fundamental multivaluedness of dynamical functions revealing, in particular, the ultimate origin of randomness in the world which does not depend on the details of the current "theory of everything". An illustration to these statements, and another fundamentally important and 'hard' case of chaos extending, in particular, the existing theory of everything, is provided by the causal description of the irreducible quantum indeterminacy of the measurement process, part IV.

It is not surprising that this extended character of the generalised method of effective dynamical functions and the ensuing concept of the fundamental dynamic uncertainty leads also to unambiguous and universal definitions for such basic notions as (non)integrability, general solution (completeness), and complexity of dynamical systems. Although the conclusions below refer, strictly speaking, to the directly analysed case of quantum Hamiltonian systems with periodical perturbation, the extensions outlined above and the evidently universal character of these conclusions permit us to propose their formulation not depending on the type of a chaotic (complex) system considered.

As concerns *nonintegrability* (or the close notion of non-separability), one easily finds that within our approach it is nothing else but the same plurality of realisations. In simple terms, this plurality is the reason for which one cannot, in principle, find a *unique* solution for a nonintegrable system by the methods effectively oriented *only for this type* of solution. However, as follows from our analysis, it is sufficient to use another, eventually more general, representation of the main dynamical equation to see that a solution *always* exists for any really existing physical system, in certain *now well-defined* sense, and it can be found; simply for the "integrable" systems we have only one realisation, and then the conventional methods may be sufficient to obtain solution, whereas for the "nonintegrable" systems the number of realisations is indeed more than one, and naturally one can obtain and analyse their solutions only within the extended formalism providing explicitly the fundamental multivaluedness. This resolves the usual ambiguity of the subject when for an arbitrary system one can hardly prove that it is nonintegrable if it is the case. From the other hand, if the system is integrable there is neither general methods to prove it and find the solution; the particular methods used, being sometimes very powerful, are never universal; typically one cannot even say how much remains beyond their applicability, and hence any systematisation is impossible. From this point of view, the concept and the formalism of the FMDF can be regarded as a useful and *universal* tool for unambiguous detection of the system (non)integrability which can then be used for finding and analysis of solutions, not only for nonintegrable but also for those integrable systems, where the known particular methods fail. This formalism represents thus a *universal non-perturbative method* of analysis applicable both to integrable and nonintegrable systems and marking the *fundamental escape from the closed space of the perturbation theory*. It explains why all those perturbation expansions and the accompanying singularities in so many different theories (or at least most of them) 'did not want' to converge: they were condemned to this because of the starting limitation, explicit or implicit, to the single-valued, non-complex dynamics; the 'singularity' of a chaotic (complex) system is its dynamic multivaluedness, and it is too sharp to be compatible with an ordinary expansion into a series. Moreover, in many cases where the perturbation series seem to converge they represent, nevertheless, only one possibility among others which exist but are lost; in this way one gets into a specific 'trap of self-consistency' of the logic of reduction. Now that the malediction of the perturbation paradigm is removed one may imagine, practically, a combined approach when the fact of integrability or nonintegrability is obtained by the FMDF formalism as well as a number of solution properties, whereas some other details can be specified by other (e. g. known) methods applied to individual realisations and using the results of the previous general analysis.

  We see now that the two couples of notions, integrability-nonintegrability and regularity-chaos, are, in fact, very close to each other. Dynamical chaos can be regarded as the nonintegrability plus the possibility of transitions between realisations assisted by additional external perturbation or noise. In cases when the latter is relatively small, randomness can be difficult to observe directly within one experimental run, while the nonintegrability should still manifest itself through the plurality of realisations observed, for example, in many equivalent runs. One may say that the notion of nonintegrability introduced above concerns just the 'dynamical', regular, part of dynamical chaos, i. e. the realisations themselves which are not random, whereas its 'chaotical', random part comes simply from the plurality of realisations assisted, to a varying degree, by noise. In particular, when the latter is sufficiently large (or the realisations are sufficiently close) we arrive at the pronouncedly randomised dynamics. In its turn, integrability means, according to our definition, the existence of only one realisation, and hence it imposes regularity. The reverse is less definite: we have seen above that in some generally regular domains chaos, and hence partial nonintegrability, can exist in an asymptotically weak form.

  Another implication of the relation plurality-nonintegrability concerns the presentation (16.II) of the *general solution* of a problem which also seems to have a universal form for all kinds of dynamical systems. In fact, it extends the notion of the general solution for the linear systems presented as a linear combination of the proper solutions from their complete set. In the general case of nonintegrable system the role of such general solution is played by the complete set of the 'proper realisations'. Now, however, they should be superimposed in the form of the probabilistic, incoherent sum though with the predetermined coefficients-probabilities. Again one can see both the external analogy, and the basic difference between the possibility, exemplified by our method, of finding an accurate, in principle, solution even for these nonintegrable systems and the integrability in the full (and now precise) sense.

There is another general notion from the same series, often applied to chaotic systems with complex behaviour, that of *nonlinearity*, and one may wonder how it is related to the above logical scheme of FMDF. Using our results, we can argue that this ordinary notion of nonlinearity based rather on mathematical properties, and above all on the violation of the linear superposition rule, plays only a supplementary, though indispensable, role in the complex system behaviour. For example, we have seen in section 3 that this ordinary nonlinearity of the unperturbed motion exerts essential amplifying effect on the system chaoticity by diminishing the effective realisation separation. It is quite clear, however, that this conventional nonlinearity is not at the very origin of complex behaviour. This conclusion is supported also by the existence of a number of the well-known exact nonlinear solutions of the nonlinear equations, like e. g. the famous solitons, which are absolutely predictable and thus not complex (see below for the rigorous definition of complexity). In return, our findings within the FMDF concept permit us to introduce the notion of the *effective nonlinearity* characterising just the complex behaviour. The ordinary nonlinearity can be associated with the existence of certain 'nonlinear' interactions in the system leading to violation of the superposition principle, etc. We can define the effective nonlinearity also as the existence of interactions, those between the elementary 'linear' modes of the integrable components (that can be, in fact, nonlinear in the ordinary sense, like e. g. solitons), which leads, as it was explained in section 5, to the excessive number of their combinations (ordinary complete solutions) equivalent to chaotic behaviour of the competing realisations.[*] Within this understanding the (effective) nonlinearity is largely equivalent to nonintegrability, chaoticity, and complexity, and can be employed rather as their synonym referring to the particular role of the internal system mode interaction in the development of its complex behaviour. In order to determine whether a system is effectively nonlinear, one should evidently perform the same decomposition in the complete realisation set and then verify if their number is more than one. A more subtle relation with the ordinary nonlinearity of a problem may concern the character of distribution of realisations within their complete set: whether it is discrete or quasi-continuous, denumerable or not. We leave further clarification of these details to future studies of various types of complex systems.

---

[*] Note that each individual realisation also bears the signs of such effective nonlinearity: it is also obtained as a result of all the internal interactions between the initial 'free modes' (degrees of freedom); in fact, the set of realisations is a sort of natural 'regrouping' of those modes in the new combinations due to their interactions (cf. the introduction of phonons in the linear solid state theory; contrary to this linear analogy, however, the new combinations are more numerous for a chaotic system). Among particular 'individual' manifestations of the effective nonlinearity we may mention e. g. the 'signatures of chaos' studied in the existing descriptions of quantum chaos [13] (appearing in our approach in the modified form, see discussion in section 5) and the fundamental fractal structure of realisations (section 4).

We see thus that most essential dynamical concepts can be unambiguously and universally defined on the basis of the central idea of the effective multiform structure of being, which necessitates the same property of multivaluedness in its most complete formal description. The basic difference of the FMDF from the existing notion of multivaluedness is due to the dynamical character of the former: we find the number of realisations from the main dynamic equations, it is not predetermined, it represents a *dynamic quantity* with intuitively transparent physical meaning and not just another mathematical characteristic. This profound physical nature of the fundamental dynamic uncertainty leads, in particular, to a decisive advance in the definition of *probability*, another basic notion closely related to chaotic behaviour. The discussion around the definition of randomness (chance, hazard, irregularity, etc.) raised in connection to the problem of, and the unambiguous criterion for, the dynamical chaos (see e. g. [4]) cannot find a consistent solution within the conventional picture which eventually always appeals to quantitative *mathematical* definitions of the formal type (exponential divergence of trajectories, or the equivalent ones). The fundamental and universal enough *physical* solution to this problem is inseparably coupled to the physical definition of probability. In the existing formal definition probability is defined as a measure on the "space of events", or the "ensemble of systems". As neither events, nor their ensemble were provided with any clear definition, all that remained was simply to count experimentally the features that had a look of the proper events, hoping not to miss anything and using empirical rules such as the "law of large numbers". Now that we have the precise definition, and the analytical method of determination, of the complete ensemble (space) of our realisations (events), eq. (12.I), the probability, introduced as the measure on this *practically obtained* space, is defined without any ambiguity and in particular can be found theoretically (in principle, analytically), before any experimental verification (see eqs. (16.II)). This is possible also due to the *natural* appearance of realisations as the irreducible multiplicity of solutions of the modified dynamic equations: they are thus *equally* probable, which provides the measure. It is at this point that we touch the profound physical roots of this new concept of probability which imparts a concrete *causal* meaning to the notion of randomness.

But what is even more important, the implication of the probabilistic set of realisations permits one to introduce the precise and basically simple measure of complexity, $C$, of a physical system as a function $C(N_\Re)$ such that:

$$C = 0 \text{ if } N_\Re = 1, \ C > 0 \text{ if } N_\Re > 1, \text{ and } dC/dN_\Re > 0 ; \tag{34a}$$

for example, for a system with dynamics $\mathcal{D}$ determined by eq. (17), one may put

$$C = C(\mathcal{D}) = f(\mathcal{D})\ln(N_\Re) , \tag{34b}$$

with $f(\mathcal{D})$ specified by some other, more particular, considerations and playing a minor role (typically, it can be just a numerical coefficient). In this definition, complexity is deduced directly from the dynamics of the system by the well-defined procedure and possesses automatically all the necessary properties.[*]) In particular, complexity defined by eqs. (34) is zero both for the absolutely regular and completely randomised dynamics (in the two cases evidently $N_\Re = 1$), and thus one can naturally satisfy this elementary condition presenting a typical difficulty for the definition of physical complexity [17].

This feature is a direct consequence of a more general and very profound physical property that should be expected, in fact, for any consistent description of complexity: the latter implies basically the existence of the irreducible, and subtle, combination of two antithetical primary principles, order (regularity) and disorder (randomness), which form a kind of natural interlacing taking a multitude of fancy forms. The properly understood notion of *chaos* means just this specific, and nonetheless dominantly spread, mixture and not simply 'something random'; it is the general name for the phenomenon realising the general property of complexity. This gives a precise and profoundly consistent meaning to the apparently paradoxical phrase, 'the order of the universe is chaos'. In our concept of the fundamental dynamic uncertainty, culminating in the definition (34), we obtain this particular property, and this is the true reason why the proposed formalism of the fundamental multivaluedness easily overcomes the 'inherent' regularity of quantum mechanics and extends the correspondence principle to arbitrary complex systems. The regularity is represented by each of the (complete) realisations and the randomness is provided by their plurality. This latter postulate is especially important: it 'explains' randomness in general, in terms of regular notions, more obvious for our basically rational consciousness; the ultimate *origin* of randomness is presented as a *law* which is quite *deterministic* as such (cf. the above-mentioned extension of the notion of probability). This removes the eternal malediction of chance in the universal harmony of Nature: chaos is indeed a form of order, a more complex one (in parts IV,V we shall see how the same concept can be used, not surprisingly, to resolve Einstein's opposition to the probabilism of the conventional quantum theory). That is another, general, reason why particular formulations of our main concept for other cases of complex behaviour can be obtained, in principle, without difficulty (we leave it to the subsequent publications).

Randomness can now be universally and unambiguously defined by the condition $C > 0$ ($N_\Re > 1$) and rigorously tested for any dynamical system using the method of FMDF. We see, however, that in reality this causal randomness never appears in 'pure' form; for any nontrivial dynamical system it is accompanied by the regularity existing within each of the realisations. That is why it is better to speak about 'chaoticity', or 'complexity', that includes randomness, or unpredictability, as an irreducible, and causally obtained, component.

---

[*]) The presence of the logarithmic function in eq. (34b) corresponds well to the hierarchical multiplicative breeding of realisations in more involved systems with several levels of dynamical complexity.

The proposed definition of complexity is as much consistent and useful in practical applications for real complex system analysis. Thus, in a straightforward fashion can one obtain from eqs. (34) the dependence of $C$ on any dynamical parameter for a system, like e. g. the notorious chaoticity parameter K for the periodically perturbed quantum Hamiltonian system: $C(K) = f(K)\ln[N_\Re(K)]$, where some most essential features of the dependence $N_\Re(K)$ have been studied above (section 3.II), and they can be further specified if necessary. We can say now, for example, that the global behaviour complexity for the time-independent system with periodic perturbation, $C_g(K)$, takes the well-defined finite values, depending on $K$, within the interval $K_c < K < K_q$, whereas outside this interval it becomes practically zero, which means that the system global dynamics is regular. Note that this conclusion, as well as its eventual refinements concerning particular dynamic regimes described in section 3.II, is obtained analytically within quantum-mechanical formalism permitting also the straightforward semiclassical transition. Performing that transition, we obtain the analogous conclusion for the corresponding classical systems, where now $K_c = 1$ and $K_q = \infty$ (for time-dependent systems the latter equality holds in the general case). This conclusion is evidently in agreement with the well-known results obtained in classical mechanics [4], but now it is expressed through the unambiguously defined physical complexity.

One of the important general properties of the complexity thus introduced is that it is completely determined by the dynamics $\mathcal{D}$ of a system and characterises its most significant features. This means that $C(\mathcal{D})$ represents one of the essential dynamical functions of a system such as energy or momentum for a mechanical system. Correspondingly, there should exist the 'complexity conservation law' stating that $C(\mathcal{D})$ of a 'complexity-conservative' dynamical system remains constant whatever the other changes during the evolution of that system. The class of the relevant systems is larger than that of the closed systems: one can well imagine a system exchanging of energy or momentum with the surroundings but preserving its complexity. For this, as is clear from eqs. (34), the dynamics of the system $\mathcal{D}$, eq. (17.II), should remain unchanged. Practically, the latter concerns especially the number of realisations, $N_\Re$. The modification of $N_\Re$, $\mathcal{D}$, and $C$ can be figured to occur typically as a result of some profound 'creative' (or 'destructive') changes of 'material' or 'structural' type introduced into the system during its 'essential' interaction with the external world. Can there exist a 'complexity flux' analogous to momentum and energy transfers, and how can one quantify the exchanges of complexity?, - questions for the future which now may have sense.

There is no place here to discuss the detailed relation of the physical complexity introduced above with the existing mathematical definitions of complexity. It is important to note, however, that our dynamical complexity turns to zero simultaneously with algorithmic complexity which is shown to be zero for the *ordinary formalism* of quantum mechanics (at least for certain large class of systems) [10]. It is clear also that algorithmic complexity (as well as any other reasonably defined complexity) will be non-zero for the chaotic dynamics ($N_\Re > 1$) described by the *modified quantum formalism* above. To understand these and other relations with formal definitions of complexity operating with "symbols", or "letters", that form "sequences", etc., it is convenient to assume that our realisations play effectively the role of such elementary symbols of dynamics. As we have completely random sequence of realisations (see eq. (16a.II)) 'chosen' by chaotic system from the dynamically determined regular "alphabet", eq. (12.I), the complexity of this sequence should be non-zero and is, in fact, close to its maximum. Such symbolisation at a much higher level of realisations avoids both physical ambiguity and formal sophistication, the characteristic difficulties of the direct low-level symbolisation of dynamics represented by discrete-time maps [18]. The high-level symbolisation, using the once found alphabet of realisations, will ensure also crucial gains in computer simulations of the natural complex system behaviour. This is not surprising: it is also the principle of functioning of those systems, and in particular, of the human brain.

These advantages reflect the basic physical origin of our symbols-realisations: they are simply the natural spatio-temporal integral elements of all hierarchical levels of being, described as *events* (temporal aspect) or *nonlinear self-organised structures* (spatial aspect) [12], that form a sequence called Life which is produced by a (non-Turing) computer called Nature.[*] The generality of this conclusion is supported by the revealed universality of our basic results: the self-consistently derived set of events-realisations with the accompanying notions of dynamics, eq. (17.II), and complexity, eqs. (34), show no limitations either in the type of a dynamic system (conservative, dissipative, open, closed, continuous, discrete, time-dependent, or -independent), or in the considered particular aspect of its behaviour (self-organisation, chaos, non-equilibrium phase transitions, etc.). We continue to specify this universality in the forthcoming publications (see also the end of section 10.V).

Note finally that our complexity extends also the usually proposed versions of entropy (see e. g. [4]) which now can be understood as a really *dynamic* measure of chaoticity (i. e. complexity) and not of specifically defined mixing, or disorder, or informational properties. In particular, if we apply the well-known Shannon definition of entropy, related to signal processing but also extensively used e. g. in statistical physics,

$$S = -k \sum_i p_i \ln p_i ,$$

to the set of realisations of a complex system, interpreting $p_i$ as the probability for a system to fall into the $i$-th realisation, $p_i = \alpha_i = 1/N_\Re$ (see eqs. (16.I)), then we easily obtain

$$S = kC/f(\mathcal{D}) ,$$

or simply

$$S = C ,$$

---

[*] The Programmer of this computer should also exist — this is a rigorous consequence of the complexity conservation law within the 'universal science of complexity' [19] (subsection 13.7).

with $f(\mathcal{D}) = k$, where $C$ is defined by (34b). Note that this definition of entropy is as universal as that of complexity, which means that it can be equally well applied to equilibrium and non-equilibrium systems containing many or quite a few degrees of freedom.[#] This provides, in particular, the natural transition from dynamical to statistical properties and the corresponding fundamental origin of irreversibility, both in dynamics and thermodynamics. It is not difficult to see that all this is possible because of the proper introduction of the 'states' (realisations) participating in the definition of entropy, together with the method of their practical obtaining and analysis (cf. the related substantiation of the extended notions of probability and randomness, above in this section).

---

[#] The formulas above are valid, strictly speaking, only for the systems with one level of complexity like e. g. periodically perturbed Hamiltonian systems analysed in detail in this work. Their behaviour is characterised by the constant value of entropy-complexity, $S = \mathrm{const}$, all the parameters being fixed. The most general and consistent definition of entropy inevitably involves the introduction of an accompanying dual notion, equally universal, that of information. We present it within the nearest future development of our concept (see [19])..

# 7. Conclusions: Quantum mechanics with chaos

A new method for the dynamical chaos analysis has been presented by its detailed application to the description of quantum chaos in periodically perturbed Hamiltonian dynamical systems. We consider in parallel two versions of such a system: the time-independent conservative system consisting of a regular part and a periodic perturbation, and the time-dependent case with a regular part disturbed by a time-periodic potential. The results for the two versions are generally quite similar with the distinctions specified where necessary.

The key result of the method consists in the discovered additional plurality of solutions of a dynamical equation (here the Schrödinger equation) presented in certain modified form, where the ordinary main dynamical function (here the potential) is replaced by the corresponding effective dynamical function. This modified form is obtained from the ordinary form by simple algebraic transformations, familiar from the optical potential formalism, which should give, a priori, the equivalent equation. However, the effective potential is found to be a multivalued function and the excessive number of solutions cannot be reduced to any known, or spurious, effect; it forms a set of 'realisations' depending on the type of a system and its parameters. Each realisation corresponds to the 'normal' complete set of solutions obtained for the corresponding 'branch' of the effective dynamical function and providing a possible value of each observable quantity, eq. (12.I). It is shown that if one accepts this modified form of the Schrödinger equation as the basic one, then the discovered existence of more than one realisation for a chaotic system can be interpreted as 'true' quantum chaos, eqs. (16.II).

Indeed, quantum chaos described in this way involves the unreduced dynamical unpredictability as well as its transformation, in the semiclassical limit, to the corresponding results for the same system obtained within classical mechanics. This agreement with classical results includes a number of qualitative features of chaos (the existence of the regimes of global chaos and global regularity and their general character, like the asymptotically vanishing remnants of chaos in the field of global regularity) and the quantitative expression for the point of transition chaos-regularity (called here the 'classical border of chaos', for the general quantum case), eqs. (18.II)-(21.II).

At the same time a number of purely quantum features is also obtained within the proposed description. One of them is the 'quantum suppression of chaos' which, contrary to the majority of the existing conceptions of quantum chaos, is generally only partial in our description, allowing of existence of the developed quantum chaos even far from the semiclassical limit. There are at least two common types of this quantum suppression of chaos.

The first one has a quite universal character and is due simply to the finite energy-level, and therefore realisation, separation in quantum case, which leads only to a partial disappearance of the manifestations of chaos.

The second one is more specific: it is observed in the case of Hamiltonian system with time-independent periodic perturbation and is absent in the time-dependent case. It is shown that this kind of quantum suppression of chaos can be either partial, or complete depending on parameters. It becomes complete, i. e. a generally chaotic system transforms into a strictly regular one, when the full energy of the system is less than certain characteristic quantity called the 'quantum border of chaos' which is close to the lowest energy level, eq. (22.II).

This conclusion has the important physical consequences because it explains, in a remarkably simple way, the absolute stability of the elementary constituents of matter like atoms and nuclei, in their ground states, despite the existence of the developed quantum chaos regimes predicted above. However, the latter can also play an important role in the behaviour of the elementary quantum bound systems. It is clear, for example, that any excited state for such system is generically chaotic. Moreover, the existence of the specific chaotic ground states can be also expected.

It is shown thus that the quantum and classical borders of chaos and the dynamical regimes that they determine represent certain rather universal types of the chaotic system behaviour. It is important to emphasize, however, that according to the performed analysis, quantum chaos is compatible with the existence of many more general and particular variants of chaotic behaviour; they are determined by the set of the possible realisations for a chaotic system and their parameter dependence illustrated in this paper. The proposed formalism permits one to reveal and study in detail these dynamic regimes for any particular quantum system.

One of the quantum chaotic regimes is especially interesting: it represents a kind of quantum chaos without any classical analogue. It appears in the form of large irregular quantum 'jumps' of the effective potential characteristics and, correspondingly, of the measured quantities. In particular, the height of a potential barrier can suddenly and considerably diminish; a barrier may effectively disappear at all, or it can even be transformed into a well, and vice versa. In the suitable experimental conditions this phenomenon may manifest itself as the specific 'chaotic quantum tunnelling'.

The same method leads to the prediction of chaotic behaviour also for the free-motion states of quantum system possessing a transparent interpretation and a classical analogue.

Another feature of the chaotic quantum dynamics revealed within the formalism of the fundamental multivaluedness is its fractal character. The main dynamical function, the effective potential, is composed of many fractal branches, each of them corresponding to a realisation of a problem. The fractal smearing is superimposed on the fundamental multivaluedness and nonlocality of the effective dynamical function that is also referred to as the fundamental dynamical fractal of a problem. It is shown how the details of the chaotic dynamics can be translated in terms of the fractal geometry of this peculiar object, which outlines the way for the complete geometrisation of the complex (quantum) dynamics.

Generalising all the results above, we may say that they form a starting point of what could be called the 'quantum mechanics of chaotic systems', which is fundamentally different from the approach of canonical 'quantum chaos' whatever is the exact meaning of the latter. Indeed, we have seen that one does need to use another, extended form of quantum formalism, eqs. (33), for a non-contradictory description of complex system behaviour, and therefore chaoticity in quantum world is something more profound than another specific physical phenomenon; in fact, it cannot be separated from the fundamentals of quantum dynamics, but at the same time there is no any direct involvement with the primary quantum postulates. In other words, the revealed fundamental dynamic uncertainty is at least as 'big' as the quantum-mechanical indeterminacy (in parts IV,V we show that the latter can, in fact, be reduced to the former). The logically natural construction of the modified quantum formalism of fundamental multivaluedness provides just a proper combination of the two concepts leading, as we have seen, to agreement with classical mechanics and to a variety of complex (in a good sense!) patterns of quantum behaviour. Further understanding and detailed investigation of this complex quantum world may constitute a subject of the new, 'chaotic' quantum mechanics.

It is also demonstrated in the present work that the proposed method and the concept of the fundamental multivaluedness of dynamical functions can be, most probably, extended to other types of chaotic systems providing thus a universal basis for the complete description of their dynamics, eq. (17.I), and the ultimate origin of randomness and complexity. In particular, an invariant definition of physical complexity is proposed, eqs. (34). It is based on the set of realisations for a system, naturally satisfies the general requirements for such definition and can be directly applied to real dynamical systems providing e. g. the dependence of their complexity on parameters. This universal concept of dynamic complexity involves also the rigorous transparent definitions, and extensions to arbitrary systems, of such notions as (non)integrability, effective nonlinearity, general solution (complete system of solutions), and probability (randomness).

The fundamental alternative concerning the complexity of the world is solved thus in favour of complexity, within our approach. Nonetheless the other possibility, outlined in the Introduction, also remains formally non-contradictory, even though to our opinion a number of Gedanken experiments could provide serious doubts in the viability of the zero-complexity world. The results of parts I-III of this work are effectively reduced to the discovery of a plausible modification of quantum mechanics with non-zero complexity, which provides effective support for these doubts about the absolutely calculable universe. Apart from doubts and hopes, however, neither of the two possibilities can be conclusively accepted or rejected at the present moment. Their further comparison needs additional theoretical and experimental work.

# References

[1]  P.H. Dederichs, Solid State Phys. **27**, 136 (1972).

[2]  G.M. Zaslavsky, *Chaos in dynamical systems*  (Harwood Academic Publishers, London, 1985).

[3]  G.M. Zaslavsky, R.Z. Sagdeev, D.A. Usikov and A.A. Chernikov, *Weak chaos and quasi-regular patterns*  (Cambridge Univ. Press, Cambridge, 1991).

[4]  A.J. Lichtenbegr and M.A. Lieberman, *Regular and Stochastic Motion* (Springer-Verlag, New-York, 1983).

[5]  E. Ott, *Chaos in dynamical systems*  (Cambridge Univ. Press, Cambridge, 1993).

[6]  H.-O. Peintgen, H. Jürgens, and D. Saupe, *Chaos and Fractals . New Frontiers of Science*  (Springer-Verlag, New-York, 1992).

[7]  S.R. Jain, Phys. Rev. Lett. **70**, 3553 (1993); I. Guarneri and G. Mantica, Phys. Rev. Lett. **73**, 3379 (1994).

[8]  J. Feder, *Fractals*  (Plenum Press, New York, 1988).

[9]  A.P. Kirilyuk, Nuclear Instrum. Meth. **B69**, 200 (1992).

[10] J. Ford and G. Mantica, Am. J. Phys. **60**, 1086 (1992); see also J. Ford, G. Mantica, and G. H. Ristow, Physica D **50**, 493 (1991); J. Ford and M. Ilg, Phys. Rev. A **45**, 6165 (1992).

[11] M.V. Berry, in *Chaos and Quantum Physics*, edited by M.J. Giannoni, A. Voros, and J. Zinn-Justin (North-Holland, Amsterdam, 1991).

[12] I. Prigogine, *From Being to Becoming*  (Freeman, San Francisco, 1980); Can. J. Phys. **68**, 670 (1990); G. Nicolis and I. Prigogine, *Exploring Complexity*  (Freeman, San Francisco, 1989).
H. Haken, *Advanced Synergetics* (Springer, Berlin, 1983).

[13] F. Haake, *Quantum signatures of chaos*  (Springer, Berlin, 1991).

[14] G. Jona-Lasinio, C. Presilla, and F. Capasso, Phys. Rev. Lett. **68**, 2269 (1992).

[15] G. Hose and H.S. Taylor, J. Chem. Phys. **76**, 5356 (1982).

[16] G. Hose, H.S. Taylor, and A. Tip, J. Phys. A **17**, 1203 (1984).

[17] G. D'Alessandro, A. Politi, Phys. Rev. Lett. **64**, 1609 (1990).

[18] A. Crisanti, M. Falcioni, G. Mantica, and A. Vulpiani, Phys. Rev. Lett. **50**, 1959 (1994).

[19] A.P. Kirilyuk, *Universal Concept of Complexity by the Dynamic Redundance Paradigm: Causal Randomness, Complete Wave Mechanics, and the Ultimate Unification of Knowledge* (Naukova Dumka, Kiev, 1997), in English. For a non-technical review see also: e-print physics/9806002.